\documentclass[aps,prc,twocolumn,showpacs,amsmath,amssymb,floatfix,footinbib]{revtex4}
\usepackage{graphicx}
\usepackage{epsfig}
\usepackage{amsmath}
\begin{document}

\title{Test of modified BCS model at finite temperature}

\author{V.~Yu.~Ponomarev$^{1,2}$ and A.~I.~Vdovin$^{1}$}
\affiliation{
$^1$Bogoliubov Laboratory of Theoretical Physics,
Joint Institute for Nuclear Research,
141980 Dubna, Russia \\
$^2$Institut f\"ur Kernphysik, Technische Universit\"at Darmstadt,
D--64289  Darmstadt, Germany}
\date{\today}

\begin{abstract}
A recently suggested modified BCS (MBCS) model has been studied at finite
temperature. We show that this approach does not allow the
existence of the normal (non-superfluid) phase at any finite
temperature. Other MBCS predictions such as a negative pairing gap,
pairing induced by heating in closed-shell nuclei, and
``superfluid -- super-superfluid'' phase transition are discussed also.
The MBCS model is tested by comparing with exact solutions for the 
picket fence model. Here, severe violation of the internal symmetry of the
problem is detected. The MBCS equations are found to be inconsistent.
The limit of the MBCS applicability has been determined to be 
far below the ``superfluid -- normal'' phase transition of the conventional 
FT-BCS, where the model performs worse than the FT-BCS.
\end{abstract}


\maketitle

\section{Motivation}

Interest in nuclear pairing correlations has been
intensified for many reasons in recent years (see, e.g., reviews
\cite{Dean03,Zel03}). Among the many aspects of the problem, a thermal
behavior of the pairing correlations in nuclei is considered also.
It is well known that the conventional thermal BCS approach
produces not very precise results when applied to finite
many-particle systems like atomic nuclei, principally due to 
particle number fluctuations.
In order to overcome, at least partially, the shortcomings of the BCS 
approach a new model, named the modified BCS (MBCS),  was suggested and 
explored in papers \cite{DZ01,DA03a,DA03b,DA03c,DA03d}. According to the 
MBCS calculations a sharp ``superfluid -- normal'' phase
transition, which is a distinct feature of the conventional thermal
BCS theory, appears at much higher temperatures and may be
smeared out completely.
In this paper we analyze the performance of the MBCS.

\section{Introduction to the MBCS model}

The conventional BCS theory is based on the Bogoliubov transformation
from particle creation $a^{\dagger}_{jm}$ and annihilation $a_{jm}$
operators to quasiparticle operators
$\{\alpha^{\dagger}_{jm},\ \alpha_{jm}\}$:
\begin{equation}
a^{\dag}_{jm} = u_j \alpha^{\dag}_{jm} + v_j \alpha_{j \tilde{m}},
\label{e0}
\end{equation}
where the index $jm$ corresponds to the level of a mean field with
quantum numbers $j \equiv [n,l,j]$, projection $m$ (we consider the
spherical case) and energy $\varepsilon_j$. Tilde in Eq.~(\ref{e0}) and 
below means time reversal operation:
$\alpha_{j \tilde{m}} = (-1)^{j-m} \alpha_{j -m}$.

The BCS equations at zero temperature $T=0$ are obtained, e.g., by
minimization of the energy of the pairing Hamiltonian:
\begin{equation}
H_{\rm pair} = \sum_{jm}\varepsilon_{j}a^{\dag}_{jm}a_{jm} -
\frac{1}{4} \sum_{{jm}{j'm'}} G_{jj'}  a^{\dag}_{jm} a^{\dag}_{j\tilde{m}}
a_{j'\tilde{m}'} a_{j'm'}
\label{ham}
\end{equation}
in the ground state (treated as a quasiparticle vacuum) under the 
condition that the number of particles $N$ in the system is conserved
on average.
Let us consider the simplest case of a constant pairing matrix element
$G_{jj'} = G$.
Then the BCS equations have the form:
\begin{eqnarray}
N & = &  2 \sum_j \Omega_j v_j^2  \nonumber \\
\Delta &=& G \sum_j \Omega_j u_j v_j,
\label{BCS}
\end{eqnarray}
where
\begin{equation}
u_j = \sqrt{\frac{1}{2} \left( 1+
\frac{\varepsilon_j-\lambda}{E_j} \right)},~~~
v_j = \sqrt{\frac{1}{2} \left( 1-
\frac{\varepsilon_j-\lambda}{E_j} \right)},
\label{vu}
\end{equation}
$ \Omega_j = (j + 1/2)$ and
$E_j = \sqrt{(\varepsilon_j-\lambda)^2+\Delta^2}$ is a quasiparticle
energy.
In these equations $\Delta$ is a pairing gap and $\lambda$ is a 
chemical potential or the energy of a Fermi level.

The nuclear Hamiltonian $H'= H_{\rm pair} -\lambda \hat{N}$ can be
rewritten in terms of the quasiparticles as
\begin{equation}
H' = U + \sum_j b_j {\cal N}_j + \sum_j c_j
({\cal A}^{\dag}_j + {\cal A}_j) + H_c + H_{res},
\label{ham2}
\end{equation}
where
\[
{\cal N}_j = \sum_m \alpha^{\dagger}_{jm} \alpha_{jm},~~~
{\cal A}^{\dag}_j = \frac{1}{\sqrt{\Omega_j}} \sum_{m>0}
\alpha^{\dagger}_{jm} \alpha^{\dagger}_{j\tilde{m}}
\]
and (see e.g. Ref.~\cite{ho})
\begin{eqnarray}\label{Ubc}
U&=&2\sum_j \Omega_j (\varepsilon_j -\lambda) v^2_j - 
G (\sum_{j'} \Omega_{j'}u_{j'} v_{j'})^2 \nonumber \\
b_j& = &(\varepsilon_j - \lambda) (u^2_j - v^2_j) + 
2 G u_j v_j\sum_{j'} \Omega_{j'}u_{j'} v_{j'}\\
c_j& = &2\sqrt{\Omega_j}(\varepsilon_j - \lambda) u_j v_j  - 
G \sqrt{\Omega_j}(u^2_j - v^2_j)\sum_{j'} \Omega_{j'}u_{j'} v_{j'}.
\nonumber
\end{eqnarray}
In Eqs.~(\ref{Ubc}) the terms which renormalize single-particle 
energies ($\sim G v_j^2$) are omitted.

When the pairing strength $G$ is weak, the BCS equations yield the
trivial solution (normal phase): $\{u_j,~v_j\} = \{0(1),1(0)\}$ for 
all $j$, i.e. $\Delta = 0$.
Above some critical value $G_{\rm cr}$, a superfluid solution appears as
energetically preferable.
The value of the pairing gap $\Delta$, which receives a positive contribution
from all levels, may be considered as a measure of how strong 
pairing is in the system.
Indeed, the $u_j v_j$ combination in the second expression of Eqs.~(\ref{BCS})
indicates how far away the system is from the trivial solution.

In the conventional BCS theory at finite temperature (FT-BCS),
minimization of the pairing Hamiltonian is replaced by the
statistical average of the free energy over the grand canonical
ensemble. The FT-BCS equations read
\begin{eqnarray}
N &=& 2 \sum_j \Omega_j \bigg[(1-2n_j)v_j^2 + n_j \bigg]
\nonumber \\
\Delta &=& G \sum_j \Omega_j  (1-2 n_j) u_j v_j,
\label{BCS-T}
\end{eqnarray}
where $n_j$ are the thermal Fermi-Dirac occupation numbers for the
Bogoliubov quasiparticles
\begin{equation}
 n_j = 1 / (1+ \exp{(E_j/T)})
\label{FD}
\end{equation}
with $u_j$ and $v_j$ coefficients and energies $E_j$ having the
same form as Eqs.~(\ref{vu}) at $T=0$. However, now they are
temperature dependent through the $\Delta$ and $\lambda$ values.

The MBCS model also starts from the Hamiltonian (Eq.~(\ref{ham})) and
the canonical Bogoliubov transformation (Eq.~(\ref{e0})). New
ingredients appear at the extension of the approach to finite
temperatures.

In brief, a temperature-dependent unitary transformation to the
Bogoliubov quasiparticles $\{\alpha^{\dagger}_{jm},\
\alpha_{jm}\}$ is applied, thus transforming them into new
bar-quasiparticles $\{\bar{\alpha}^{\dagger}_{jm},\
\bar{\alpha}_{jm}\}$:
\begin{equation}
\bar{\alpha}^{\dag}_{jm} = \sqrt{1-n_j} \alpha^{\dag}_{jm} +
 \sqrt{\vphantom{1}n_j} \alpha_{j \tilde{m}}.
\label{trans}
\end{equation}
A new ground state $|\bar{0}
\rangle$ is introduced as a vacuum for the bar-quasiparticles
\begin{equation}
\langle\bar{0} | \bar{\alpha}^{\dag}_{jm} \bar{\alpha}_{jm} |  \bar{0}
\rangle =0.
\label{vac}
\end{equation}
The coefficients in the transformation, Eq.~(\ref{trans}), are selected
so that 
\begin{equation}
\langle\bar{0} | \alpha^{\dag}_{jm} \alpha_{jm} |  \bar{0}
\rangle = n_j
\label{e10}
\end{equation}
and it is assumed that the occupation numbers $n_j$ for the Bogoliubov 
quasiparticles should have the same form, Eq.~(\ref{FD}), as in the 
statistical approach.

Combining Eq.~(\ref{e0}) and Eq.~(\ref{trans}) the particle
operators $\{a^{\dagger}_{jm},\ a_{jm}\}$ are expressed in terms
of the bar-quasiparticle $\{\bar{\alpha}^{\dagger}_{jm},\
\bar{\alpha}_{jm}\}$ ones
\begin{equation}
a^{\dag}_{jm} = \bar{u}_j \bar{\alpha}^{\dag}_{jm} +
\bar{v}_j \bar{\alpha}_{j \tilde{m}},
\label{e5}
\end{equation}
where
\begin{equation}
{\bar u}_j = u_j \sqrt{1-n_j} + v_j \sqrt{\vphantom{1}n_j},~~~
{\bar v}_j = v_j \sqrt{1-n_j} - u_j \sqrt{\vphantom{1}n_j}.
\label{uv}
\end{equation}

Since the expectation value of $H_{\rm pair}$ at $T \ne 0$ in the
$|{\bar 0} \rangle$ ground state looks similar in terms of
$\bar{u}_j$ and $\bar{v}_j$ coefficients to the one at zero
temperature in terms of $u_j$ and $v_j$, the MBCS equations are
written down in analogy with Eqs.~(\ref{BCS}) as
\begin{eqnarray}
N & = &  2 \sum_j \Omega_j \bar{v}_j^2  \nonumber \\
\bar{\Delta} &=& G \sum_j \Omega_j \bar{u}_j \bar{v}_j
\label{MBCS}
\end{eqnarray}
or in terms of $u_j$ and $v_j$ coefficients and thermal
quasiparticle occupation numbers $n_j$
\begin{eqnarray}
N &=& 2 \sum_j \Omega_j \bigg[(1-2n_j)v_j^2 + n_j
- 2\sqrt{n_j(1-n_j)}u_jv_j\bigg]
\nonumber \\
{\bar \Delta} &=&
G \sum_j \Omega_j \bigg[(1-2n_j)u_j v_j
 -   \sqrt{n_j (1-n_j)} (u_j^2 -v_j^2) \bigg].
\nonumber \\
\label{MBCS2}
\end{eqnarray}

\section{Thermal behavior of the MBCS pairing gap}

In Refs.~\cite{DZ01,DA03a,DA03b}, applying the MBCS to study the thermal
behavior of different nuclear quantities, the authors point out the
following distinctive features of the new model: a) the
pairing gap decreases monotonically as temperature increases and does
not vanish even at very high $T$; b) the ``superfluid -- normal''
phase transition is completely washed out.

Taking the MBCS equations as they have been suggested, we
analyze the validity of the above results.
We have repeated the MBCS calculations for neutrons in Ni
isotopes and for neutrons and protons in $^{120}$Sn in Ref.~\cite{DA03a} 
and Ref.~\cite{DA03b}, respectively. The MBCS Eqs.~(\ref{MBCS2})
have been solved with an accuracy of $10^{-11}$. Our code
reproduces excellently all the results in Ref.~\cite{DA03a}. As
typical examples, we use in this presentation the nuclei $^{76}$Ni
(quasibound calculation, single particle levels in the continuum having 
no width), 
$^{84}$Ni (resonant-continuum calculation, finite width is taken into account
for the levels with positive energy) and $^{120}$Sn.

\begin{figure}
\epsfig{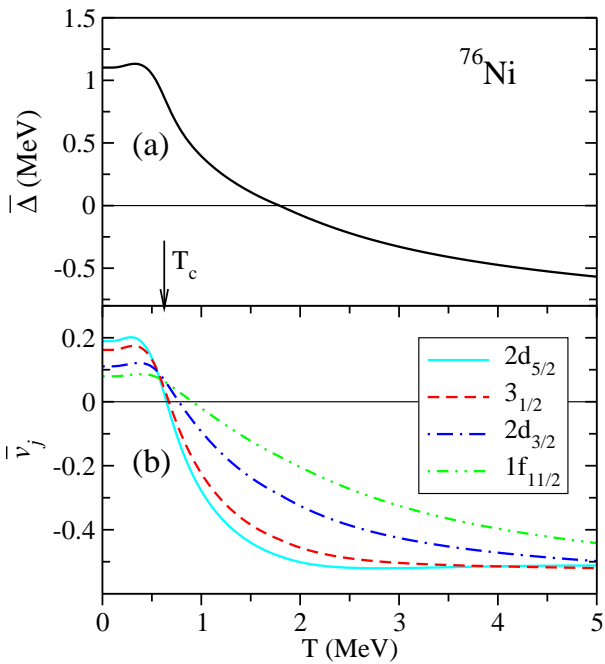}
\caption{\label{f1}
The MBCS pairing gap $\bar{\Delta}$ (upper part (a)) and
$\bar{v}_j$ coefficients (lower part (b)) for particle levels near the 
Fermi surface in $^{76}$Ni (neutrons).
The conventional critical temperature $T_c$ is shown by the vertical 
arrow and the single particle spectrum used is from Ref.~\cite{DA03a}.
}
\end{figure}

Starting with the thermal behavior of the pairing gap. The
neutron MBCS pairing gap in $^{76}$Ni is plotted in Fig.~\ref{f1}(a).
One notices that it reaches zero at $T \approx 2.1$~MeV and
continues to decrease with the negative sign. These calculations
were performed as reported in Ref.~\cite{DA03a} on a truncated
single-particle basis assuming the $N=0-28$ inert core. 
Later it was recommended in Ref.~\cite{DA03b} that
the MBCS calculations should be performed on an entire or as large 
as possible single-particle spectrum. However, though in Ni isotopes this
recipe of ``entire spectrum'' helps to avoid negative values of the 
pairing gap, it does not work in the case of $^{120}$Sn (see below).

In calculations with a wider single-particle spectrum in
Ni isotopes we have found that the gap starts to continuously
increase above a certain temperature remaining always
positive. The pairing strength has been renormalized to
keep the $\bar{\Delta}(T=0)$ value.

\begin{figure}
\epsfig{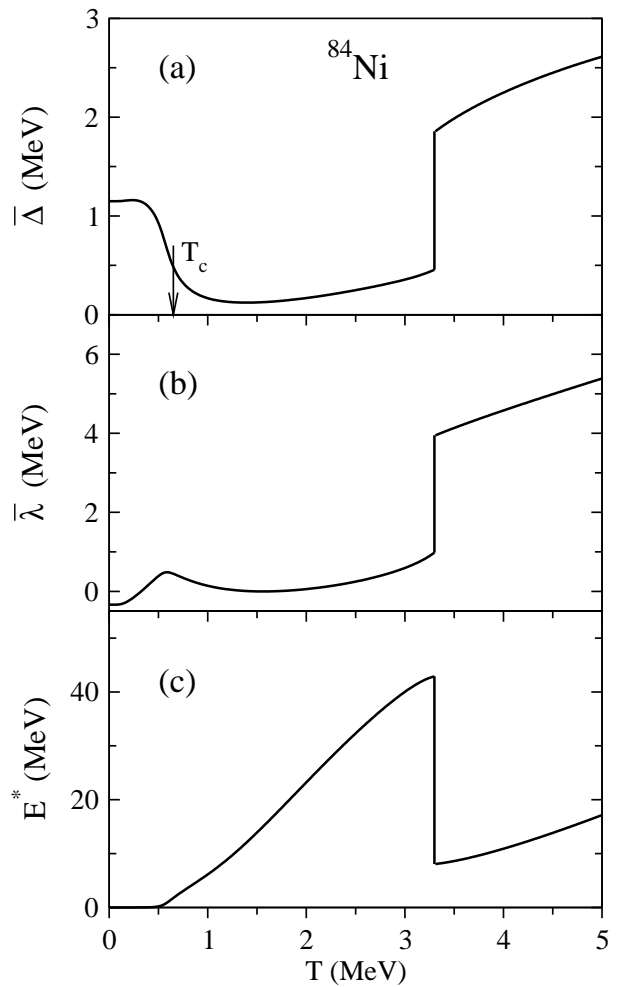}
\caption{\label{f2}
The MBCS pairing gap $\bar{\Delta}$ (upper part (a)), chemical potential 
$\bar{\lambda}$ (middle part (b)), and excitation energy ${\rm E}^*$ 
(lower part (c)) in $^{84}$Ni (neutrons).
Single particle spectrum from Ref.~\cite{DA03a} is extended by including
levels from $N=0-28$ shells.}
\end{figure}

An example of such behavior of the pairing gap is presented in
Fig.~\ref{f2}(a) for $^{84}$Ni. As can be seen, at $T\approx 3.3$~MeV
some strange discontinuities are apparent, a phenomenon which may be
defined as a ``superfluid -- super-superfluid'' (S-SS) phase transition. 
At this temperature, the MBCS equations find a new
energetically-preferable solution. A $T$-dependence of the
total excitation energy given by
\begin{equation}
{\rm E}^* = {\cal E}(T) - {\cal E}(0),
\label{E*}
\end{equation}
where ${\cal E}(T)$ is calculated from Eq.~(45) in Ref.~\cite{DA03a}, is 
shown in Fig.~\ref{f2}(c). 
The chemical potential jumps away from the $\lambda(T=0)$ value 
at $T\approx 3.3$~MeV (see Fig.~\ref{f2}(b)).

In short, while at the $\bar{\Delta}=0$ point the behavior of 
physical observables is smooth (see Fig.~\ref{f1}(b)), 
the discontinuities at the S-SS point in excitation energy, 
pairing gap and chemical potential take place. Thus, there is no phase 
transition from the superfluid to normal phase within the MBCS but 
instead a phase transition of a new type is predicted at finite 
temperatures.

\begin{figure}
\epsfig{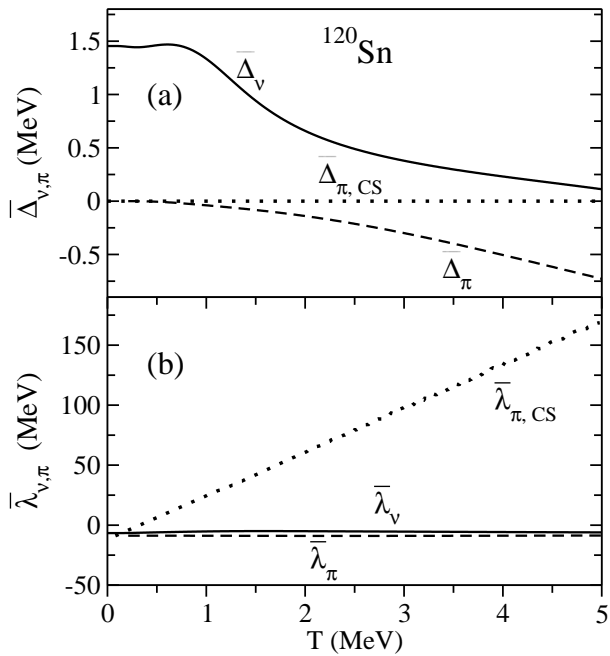}
\caption{\label{f3}
The MBCS pairing gap $\bar{\Delta}$ (upper part (a)) and chemical potential
$\bar{\lambda}$ (lower part (b)) for neutrons (solid lines) and protons 
(dashed lines) in $^{120}$Sn.
$\bar{\Delta}_{\rm CS}$ and $\bar{\lambda}_{\rm CS}$ (dotted lines) 
correspond to closed-shell calculations in Ref.~\cite{DA03b}.
See text for details.
}
\end{figure}

The absence of the normal phase in all MBCS calculations has motivated us 
to apply this model to a magic-number
system of nucleons in which the existence of this phase at
$T=0$ is expected. The nucleus $^{120}$Sn has been taken as
an example. The MBCS neutron pairing gap (solid line in
Fig.~\ref{f3}(a)) in this nucleus shows the same behavior as
already discussed for $^{76}$Ni even though a rather complete single
particle basis has been employed in $^{120}$Sn.
The neutron pairing gap vanishes at $T\approx5.5$~MeV and becomes 
negative at higher temperatures.

The proton gap in $^{120}$Sn exhibits a rather strange behavior
as a function of $T$. Starting from zero value at $T=0$ the
gap smoothly develops to a value of $-0.73$~MeV at $T=5$~MeV (dashed
line in Fig.~\ref{f3}(a)). 
A completely different result for the MBCS proton pairing gap in this
nucleus has been reported in Ref.~\cite{DA03b} and it is shown as the 
dotted line $\bar{\Delta}_{\rm CS}$ in Fig.~\ref{f3}(a). 
To obtain it, it has been suggested that closed-shell systems should 
be treated differently from open-shell ones, namely, all summations in 
the MBCS equations (\ref{MBCS2}) should be carried over hole levels only.  

In our opinion, the last recipe contradicts the previous
recipe of ``entire spectrum'' in the same publication.
It is also evident that this artificial constraint cannot be 
justified for a heated system. 
Indeed, both subsystems, the former hole (fully occupied) and particle 
(empty) levels, become partially occupied
due to the heating and there are no physical reasons to ignore the 
particle part of a single-particle spectrum. Moreover, to keep the 
number of nucleons $N$ constant under the above constraint, forces the MBCS
equations to abnormally renormalize the corresponding 
chemical potential. In the above CS-calculations the $\bar{\lambda}$ 
quantity runs away from $\bar{\lambda}_{\pi, {\rm CS}} (T=0) = -10$~MeV  
to $\bar{\lambda}_{\pi, {\rm CS}} (T=5) = +175$~MeV (following the dotted 
line in Fig.~\ref{f3}(b)). In addition, the capacity of the system is zero 
in the CS-approximation, i.e. the system is artificially frozen.

The above MBCS results are natural consequences of Eqs.~(\ref{MBCS2}). 
For example, these
equations do not have the trivial solution $\{{\bar u}_j,~{\bar
v}_j\} = \{0(1),1(0)\}$. Indeed, from Eqs.~(\ref{uv}) it would
correspond to
\\[1mm]
\begin{tabular}{lll}
\hspace*{6mm}
$u_j = \sqrt{1-n_j}$; & ~~~$v_j = \sqrt{\vphantom{1}n_j}$ 
&~~~-- particles \\
\hspace*{6mm} $u_j = -\sqrt{\vphantom{1}n_j}$; & ~~~$v_j =
\sqrt{1-n_j}$& ~~~-- holes
\end{tabular}
\\[1mm]
and contradict the positive definition of $u_j$.

One may also notice from Eqs.~(\ref{uv}) that ${\bar v_j}$
coefficients become negative for particle levels above a certain 
temperature, with $n_j$ increasing, since $v_j << u_j$. 
In which case  the MBCS pairing gap $\bar{\Delta}$ receives a positive 
contribution from hole levels and a negative contribution from 
particle levels. Thus, the
contribution of single-particle and single-hole states to a
pairing phenomenon appears to be essentially different. 

Numeric calculations (see Fig.~\ref{f1}(b)) show that two
terms in the second expression of Eqs.~(\ref{MBCS2}) compensate each 
other around the critical temperature of the conventional BCS $T_c
\approx 0.57 \cdot \Delta_{T=0}$ for particle levels and ${\bar
v_j}$ become negative at higher temperatures. The gap
${\bar \Delta}$ may vanish at some temperature but only when a
negative contribution from particles and positive contribution
from holes cancel each other (notice the difference with the
conventional BCS $\Delta =0$). However if this happens, at higher
$T$ the balance appears to be broken and ${\bar \Delta}$ becomes
finite again. It also means that $\bar{\Delta}=0$ has nothing
in common with the normal phase and that one cannot conclude from the 
absolute value of the pairing gap how strong the pairing is in 
the system.

The temperature behavior of the pairing gap depends on a
delicate balance between particle- and hole-parts of the 
single-particle basis used which makes the MBCS predictions 
very doubtful.

\section{MBCS and exact solutions of the pairing Hamiltonian}

In this section we compare MBCS predictions with exact solutions
of the pairing Hamiltonian employing the Picket Fence Model (PFM)
\cite{PFM} which is widely used as a test model for the pairing
problem (see, e.g., Ref.~\cite{sto}).
For numeric calculations we have selected $N=10$ levels, each of 
two-fold degenerate (for spin up and spin down), with the energy
difference of 1~MeV and 10 particles distributed over
the levels. This configuration thus represents 5 levels for holes with
energies $\varepsilon_{-i} = -0.5$~MeV, $-1.5$~MeV, etc and 5 
levels for particles with energies $\varepsilon_i=+0.5$~MeV, 
$+1.5$~MeV, etc.

\begin{figure}
\epsfig{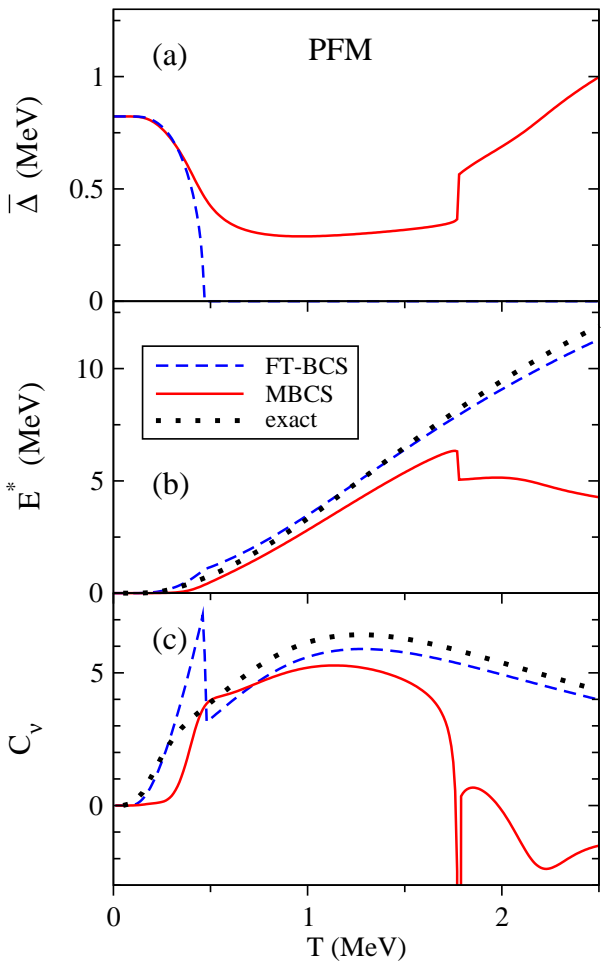}
\caption{\label{f4}
Pairing gap $\bar{\Delta}$ (upper part (a)), excitation energy 
${\rm E}^*$ (middle part (b)), and specific
heat (lower part (c)) $C_{\nu}$ as predicted by the MBCS (solid lines) 
and FT-BCS (dashed lines) for the PFM ($N=10$ and $G=0.4$~MeV).  
The exact results in (b) and (c) are plotted by dotted lines.
}
\end{figure}

The MBCS predictions for the pairing gap, excitation energy,
and specific heat given by
\[
C_{\nu} = \frac{\partial {\cal E}}{\partial T}
\]
are shown in Fig.~\ref{f4} by the solid lines. Here, also shown are 
the FT-BCS results as dashed lines (explicit expressions for the quantity 
${\cal E}$ in both approaches are given below). 
Both the MBCS and FT-BCS can be compared to the
exact solutions of the pairing Hamiltonian (dotted lines). 
It should be noted that the
MBCS-PFM pairing gap behavior is qualitatively very similar to the
one in $^{84}$Ni reported in Fig.~\ref{f2}(a) and that the ``superfluid
-- super-superfluid'' phase transition takes place at $T \approx
1.78$~MeV.

One concludes from Fig.~\ref{f4} that the MBCS does not achieve its
main goal of improving upon the description of heated nuclei in the
conventional FT-BCS. Indeed, except for a narrow region around
$T_c$, the deviation from the exact results is worse in the MBCS case
compared to the conventional FT-BCS, which persists even at a very low 
temperature.
It is also obvious that the ``superfluid -- super-superfluid'' phase
transition as well as the overall behavior at higher
temperatures are artificial effects of the MBCS.

\begin{figure}
\epsfig{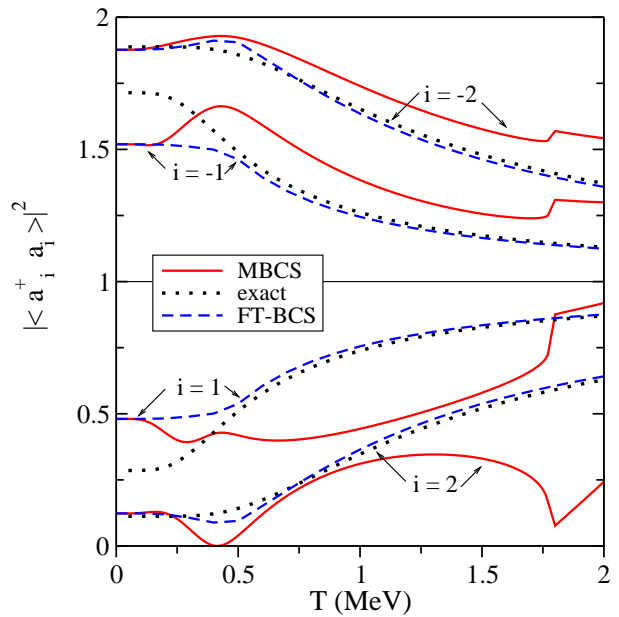}
\caption{\label{f5}
Spectroscopic factors of two lowest particle and hole levels in the PFM
($N=10$ and $G=0.4$~MeV)
calculated within the MBCS (solid lines) and FT-BCS (dashed lines).
Dotted lines represent the exact results.
}
\end{figure}

In addition, a more detailed analysis shows that the situation with
the MBCS is much worse than the disagreement apparent in Fig.~\ref{f4}. 
We further present in Fig.~\ref{f5} the spectroscopic factors for 
two particle and two hole levels closest to the Fermi surface, within the 
MBCS (solid lines) and FT-BCS (dashed lines) for a comparison with the 
exact results (dotted lines). 
It is important to keep in mind that the pairing Hamiltonian in 
the PFM possesses particle-hole symmetry. 
For this reason, dotted curves in Fig.~\ref{f5} for hole 
$-i$ and particle $i$ levels are ideally symmetric about the $y=1$ line. 
The same is true for the FT-BCS results.
Nothing of this symmetry remains after the secondary Bogoliubov 
transformation of Eq.~(\ref{trans}) is applied in the MBCS (see
solid lines in the same figure).
In addition, the description of this physical observable in the
MBCS is very poor compared to the FT-BCS results.
 
\begin{figure}
\epsfig{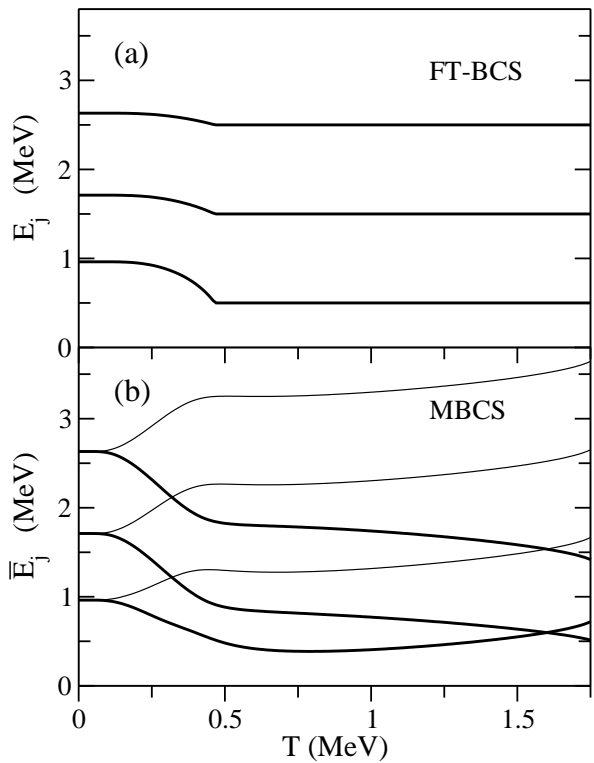}
\caption{\label{f6}
Quasiparticle spectrum in the PFM ($N=10$ and $G=0.4$~MeV)
within the FT-BCS (upper part (a)) and the MBCS (lower part (b)). 
Particle (hole) levels are plotted by thick (thin) lines.
}
\end{figure}

Breaking of the particle-hole symmetry in the MBCS calculations is even
more clearly seen from an analysis of the MBCS quasiparticle spectrum 
presented in Fig.~\ref{f6}(b).
In the PFM it should be two-fold degenerate 
(i.e. $\bar{E_i} \equiv \bar{E}_{-i}$)
as, e.g., in the FT-BCS calculation in Fig.~\ref{f6}(a) because of this 
symmetry.
The chemical potential $\bar{\lambda}$ in the MBCS calculations does not 
stay at zero energy (as it should be)
but runs away to positive values as temperature increases.
This explains why the level $i=2$ appears at a lower energy than the
$i=1$ level above $T \approx 1.6$~MeV.
It is also the origin of the $E_i/E_{-i}$ splitting in these calculations.  

We have repeated the MBCS-PFM calculations with different values of the
pairing strength $G > G_{\rm cr}$.
As $G$ increases, the MBCS ``superfluid -- super-superfluid'' phase 
transition takes place at a lower temperature and the splitting 
of $i$ and $-i$ levels becomes stronger.

\begin{figure}
\epsfig{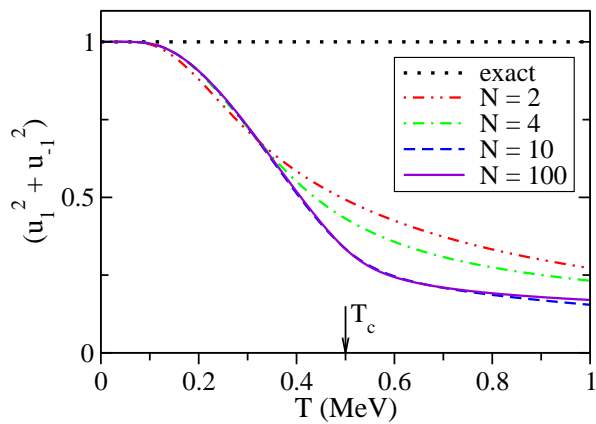}
\caption{\label{f7}
The $(u_{1}^2 + u_{-1}^2)$ quantity in the MBCS-PFM calculations
with different $N$.
$T_c=0.5$~MeV is the same in all calculations.
Analytical identity (y=1) is plotted as a dotted line.
}
\end{figure}

Another consequence of the particle-hole symmetry is that
under any conditions it should be true that
$u_i \equiv v_{-i}$ and $v_i \equiv u_{-i}$ or alternatively
\begin{equation}
u_i^2 + u_{-i}^2 \equiv u_i^2 + v_{i}^2 = 1\ . 
\label{exa}
\end{equation}
In Fig.~\ref{f7} we demonstrate what happens to this analytical identity 
in the MBCS. 
The calculations have been performed for a different number of levels $N$ 
of the PFM and
the strength parameter G has been adjusted to keep $T_c = 0.5$~MeV in 
each calculation.
The convergence of the results in this $T$-range 
(also for $\bar{\Delta}$, $\bar{\lambda}$, and $E^*$) is reached at 
$N \approx 10$.
The conclusion from Fig.~\ref{f7} is that the analytical identity of
Eq.~(\ref{exa}) is completely broken in MBCS calculations even from very 
low temperatures.

Thus, one can see that the MBCS severely violates the symmetry between
particles and holes which is an essential feature
of the pairing problem solved for the PFM. Accordingly, the MBCS can be
applied to this problem only at temperatures much lower than $T_c$.

\section{Inconsistencies of MBCS and MHFB approaches}

In this section the evaluation of the MBCS equations is analyzed to verify
consistency. To start with 
we agree that the pairing Hamiltonian $\bar{H'}$ expressed via the MBCS
variables $\bar{u}_j$ and $\bar{v}_j$  has the same form as the
BCS Hamiltonian $H'$ of Eq.~(\ref{ham2}) at $T = 0$.
It is also true that the expectation value of the pairing Hamiltonian 
$\langle \bar{0} |H_{\rm pair} |\bar{0} \rangle$  at $T \ne 0$ looks
similar to
$\langle 0 |H_{\rm pair} |0 \rangle$ in the case of the BCS at $T = 0$.
But then the formal similarity is broken and the corresponding 
energies $\bar{E}_j$ are attributed to the conventional Bogoliubov
quasiparticles and not to the ``bar''-quasiparticles.
These energies $\bar{E}_j=\sqrt{(\varepsilon_j - \bar{\lambda})^2 +
\bar{\Delta}^2}$ enter through new thermal occupation numbers $n_j$ 
and new $u,v$-coefficients
\begin{equation}\label{ren_uv}
u_j = \sqrt{\frac{1}{2} \left( 1+
\frac{\varepsilon_j-\bar{\lambda}}{\bar{E}_j} \right)},~~~
v_j=\sqrt{\frac{1}{2} \left( 1-
\frac{\varepsilon_j-\bar{\lambda}}{\bar{E}_j} \right)}
\end{equation}
(see remarks on page 8 in Ref.~\cite{DA03b} just after Eq.~(86)).

In another words, the MBCS procedure yields new eigen energies of
Bogoliubov quasiparticles while new eigen states are now modified
quasiparticles. The point here is that if one would take the mean of
the modified
quasiparticle energies under $\bar{E}_j$, they should coincide with
the BCS $E_j$ at $T=0$. Indeed, the secondary Bogoliubov
transformation of Eq.~(\ref{trans}) is a unitary one and as such,
cannot change the eigenvalues of the Hamiltonian.

There are several possibilities to derive analytically the BCS equations
from  the expectation value $\langle 0 |H_{\rm pair} |0 \rangle$ at $T = 0$.
One of them is presented, e.g., in Ref.~\cite{ho}. The BCS equations are
obtained by demanding that
\begin{equation}
 b_j\equiv E_j ~~~{\rm and}~~~c_j \equiv 0,
\label{b-c}
\end{equation}
where $b_j$ and $c_j$ are defined in Eqs.~(\ref{Ubc}).
The first expression in Eqs.~(\ref{b-c}) means that the pairing 
Hamiltonian is diagonalized in the quasiparticles space, the second one 
indicates that the so-called ``dangerous diagrams'' are excluded from 
the theory.
The solution of the BCS equations is unique. In another words, it is 
absolutely necessary that Eqs.~(\ref{b-c}) are fulfilled exactly to have the 
BCS equations in the form given in Eq.~(\ref{BCS}). 
Let us verify this for the MBCS.

\begin{figure}
\epsfig{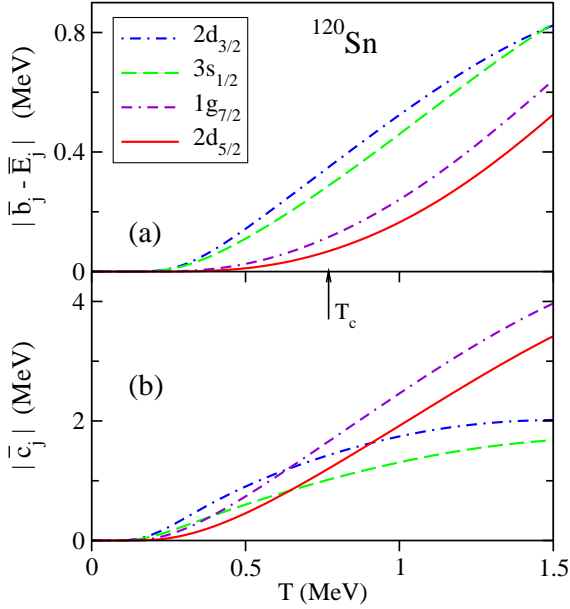}
\caption{\label{f8}
The MBCS $\bar{b}_j$ and $\bar{c}_j$ quantities as functions of temperature
for some neutron levels in $^{120}$Sn. See text for details.
}
\end{figure}

We introduce ${\bar{b}_j}$ and ${\bar{c}_j}$ quantities to replace the 
$\{u_j,v_j\}$ coefficients in Eqs.~(\ref{Ubc}) by $\{\bar{u}_j,\bar{v}_j\}$
coefficients and calculate the later from the MBCS equations.
The differences $|\bar{b}_j - \bar{E}_j|$ and ${\bar{c}_j}$ quantities
for several neutron sub-shells in $^{120}$Sn are presented in 
Fig.~\ref{f8}(a) and Fig.~\ref{f8}(b), respectively, showing that
\begin{equation}
 \bar{b}_j\neq \bar{E}_j ~~~{\rm and}~~~\bar{c}_j \neq 0~.
\label{b-c-bar}
\end{equation}
In addition, Eqs.~(\ref{b-c}) are also not fulfilled within the MBCS.

We remind a reader that the MBCS equations were not obtained analytically
but were written down in analogy with the $T=0$ BCS equations and, as far 
as the MBCS founders had noticed, using a formal similarity in some BCS 
and MBCS expressions.
But, as a matter of fact, the basic MBCS equations contained in 
Eqs.~(\ref{MBCS}) or Eqs.~(\ref{MBCS2}) 
cannot be reached from the expectation value 
$\langle \bar{0} |H_{\rm pair} |\bar{0} \rangle$ at finite $T$
because Eqs.~(\ref{b-c-bar}).

One finds in the literature~\cite{DA03b} that the MBCS equations 
may be obtained from a modified HFB model called the MHFB. 
Here, it was noticed that after application of the secondary 
Bogoliubov transformation of Eq.~(\ref{trans}), a generalized 
particle-density matrix at finite temperature ``formally looks the same 
as the usual HFB approximation at $T=0$''~\cite{DA03b}.
Then, ``following the rest of the derivation as for the 
zero-temperature case''~\cite{DA03b}, the MHFB equations were
written down.
Below we verify thermodynamical consistency of the MHFB (MBCS) since
Ref.~\cite{DA03b} lacks such an analysis.

The total energy of the system in the statistical approach has
the form
\begin{equation}
<{\rm H}> = {\rm Tr}(HD),
\label{ene1}
\end{equation}
where $D$ is a density operator and $<\ldots>$ means averaging over
the grand canonical ensemble. 
After grand potential minimization, one obtains an expression for the 
system energy as
\begin{equation}
{\cal E_{\rm BCS}} = 2 \sum_j \Omega_j \varepsilon_j 
\left[ (1-2n_j)v^2_j +n_j \right] - \Delta^2/G
\label{ene2}
\end{equation}
in the FT-HFB (or FT-BCS). 
In the MHFB (or MBCS) it has the form
\begin{eqnarray}
{\cal E_{\rm MBCS}} &=& 2 \sum_j \Omega_j \varepsilon_j 
\left[ (1-2n_j)v^2_j +n_j \right. \nonumber \\
&-& \left. 2 \sqrt{n_j(1-n_j)}~u_j v_j \right] 
- \bar{\Delta}^2/G~.
\label{ene3}
\end{eqnarray}
Equation (\ref{ene3}) appears as Eq.~(83) in 
Ref.~\cite{DA03b} or as Eq.~(45) in Ref.~\cite{DA03a}.

\begin{figure}
\epsfig{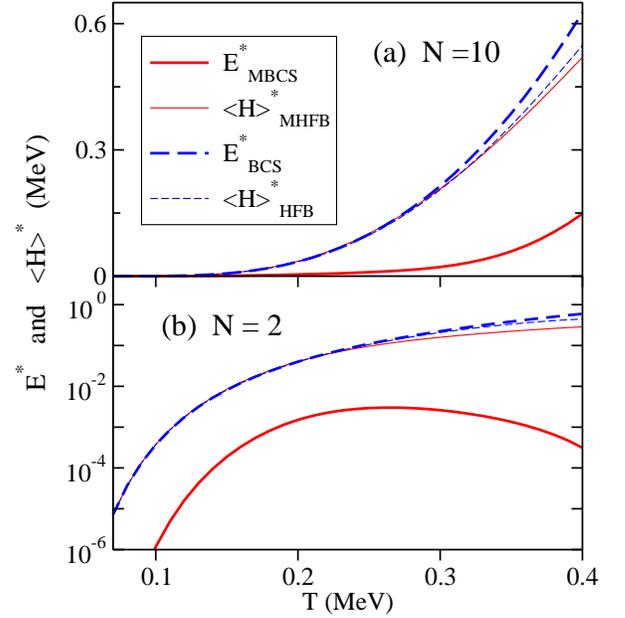}
\caption{\label{f9}
Excitation energy of the PFM calculated 
as ${\rm E}^*$ (thick lines) and  $<{\rm H}>^*$ (thin lines)
in the MHFB (solid lines) and the FT-HFB (dashed lines) with
$T_c = 0.42$~MeV for $N=10$ (upper part (a)) and $N=2$ (lower part (b)).
}
\end{figure}

In Fig.~\ref{f9}(a) we present the low-$T$ part of Fig.~\ref{f4}(b) 
(up to $T_c$) 
with predictions of the MBCS (solid lines) and FT-BCS (dashed lines) for 
the system excitation energy. 
It should be reminded that the results in Fig.~\ref{f4}(b) were 
obtained for the 
${\rm E}^*$ quantity (see Eq.~(\ref{E*})) with ${\cal E}$ calculated
in the MBCS and FT-BCS from Eq.~(\ref{ene3}) and Eq.~(\ref{ene2}),
respectively.
These results are plotted by thick lines in Fig.~\ref{f9}(a).
The excitation energy calculated in the MHFB and FT-HFB as 
$$<{\rm H}>^* = <{\rm H}>(T) - <{\rm H}>(0)$$ (in the $G_{jj'} = G$ limit
for exact comparison with the MBCS and FT-BCS results)
is shown by thin lines in the same figure.
Since $D_{\rm MHFB}$ (Eq.~(66) in Ref.~\cite{DA03b}) does not equal 
$D_{\rm HFB}$ (Eq.~(12) in Ref.~\cite{DA03b}), the $<{\rm H}>_{\rm MHFB}^*$ 
and $<{\rm H}>_{\rm HFB}^*$ quantities are slightly different in  
Fig.~\ref{f9}.

In any consistent model the values $<{\rm H}>^*$ and  ${\rm E}^*$ should not
differ significantly since they represent the same physical observable. 
One notices that the accuracy of the FT-BCS becomes worse on approaching 
$T_c$ as is well known, and disagreement reaches a few percent.
The picture within the MHFB (MBCS) is completely different: 
the ${\rm E}^*_{\rm MBCS}$ quantity is several times smaller than the
$<{\rm H}>^*_{\rm MHFB}$ quantity almost at any $T$ in this 
example~\cite{footnote}.
We have repeated the calculations in Fig.~\ref{f9}(a) also for $N=8$ and
12 keeping $T_c$ fixed. 
The differences between the $N=8,$ 10, and 12 results are hardly noticed by 
eye for each line in Fig.~\ref{f9}(a), i.e. convergence of the results 
in this example is reached with $N=8-10$.
The correspondence between $<{\rm H}>^*$ and ${\rm E}^*$ remains still
acceptable for the FT-HFB (FT-BCS) even for $N=2$ (see 
Fig.~\ref{f9}(b)).
On the other hand, disagreement in the MHFB (MBCS) results reaches several 
orders of magnitude in the $N=2$ case (notice logarithmic y-scale in 
Fig.~\ref{f9}(b)).

To conclude, thermodynamical inconsistency of the MHFB (MBCS) is obvious 
from this example. This inconsistency is detected from $T<<T_c$.

The last result clearly demonstrates that the MBCS approach is not justified 
in the framework of the usual statistical approach. This conclusion can be 
reached by other reasoning as well. The point is that the basic MBCS 
equations given by Eqs.~(\ref{MBCS2}) obtained via the secondary 
Bogoliubov transformation cannot be considered as a result of the thermal 
averaging over the grand canonical ensemble. 
Indeed, the bar-quasiparticles in Eq.~(\ref{trans}) and the new 
correlated ground state 
\[
|\bar0\rangle  =  \prod_{jm} \left( \sqrt{1-n_j} + \sqrt{n_{j}}\alpha^{+}_{jm}
\alpha^{+}_{j\tilde{m}} \right)|0\rangle
\]
are temperature dependent although the MBCS founders do not use this 
terminology. 
A possibility to deduce the basic MBCS equations via 
a variational procedure in application to the average value 
$\langle\bar{0}| H |\bar{0}\rangle$ contradicts what 
was proven long ago \cite{tak}: a ground state with 
such a property cannot be constructed in principle in the space spanned 
by eigenvectors of a quantum Hamiltonian (see also footnote on page 3
in Ref.~\cite{DA03a}).

\section{Conclusions}

In this article we have studied the modified BCS model suggested
and explored in a series of papers \cite{DZ01,DA03a,DA03b,DA03c,DA03d}. 
We have shown that this model yields many unphysical predictions:
a negative value for the pairing gap, the pairing correlations induced by
heating in the closed-shell systems, the ``superfluid -- super-superfluid''
phase transition. It also predicts that the normal phase does not
exist at any finite temperature.
The MBCS has been tested on the picket fence model for which an exact 
solution of the pairing problem is available.
In addition to rather poor description of the exact solutions, it has been 
found that the model severely violates the internal particle-hole symmetry 
of the problem.

Analysis of the MBCS equations, their derivation, and description of
different physical observables by this model have led us to the following 
general conclusion: 
The $T$-range of the MBCS applicability has been determined to be far below 
the conventional critical temperature $T_c$.
Within this narrow temperature interval the MBCS performance
is worse in comparison with that of the conventional FT-BCS.
Moreover, the MBCS is found to be thermodynamically inconsistent.

\section*{Acknowledgment}
We thank Dr. A.~Storozhenko for the exact results of the PFM
and Dr. J.~Carter for careful reading the manuscript. 
The work was partially supported by the Deutsche
Forschungsgemeinschaft (SFB 634).


\begin{thebibliography}{99}
\bibitem{footnote}
Very small, almost zero, ${\rm E}^*$ and $C_{\nu}$ values 
for $T \alt T_c$ is a distinctive feature of other MBCS predictions as 
well (see Fig.~\ref{f2} above and Fig.~4 in Ref.~\cite{DA03a}).

Rather different behavior of the $C_{\nu}(T)$ value at low $T$ in the MBCS 
is reported in Fig.~8a of Ref.~\cite{DA03b} for $^{120}$Sn. How it has 
been obtained and why it is so different from similar calculations in Ni 
isotopes in Ref.~\cite{DA03a} by the same authors, remains unclear to us.
Among what we have checked, this is the only result from all previously 
published MBCS predictions which we did not manage to reproduce.
\end{thebibliography}
\end{document}